\documentclass[useAMS,usenatbib]{mn2e}
\usepackage{graphics,epsfig}
\def \be{\begin{equation}}
\def \ee{\end{equation}}
\def \bea{\begin{eqnarray}}
\def \eea{\end{eqnarray}}
\def\etal{{et al.\ }}

\title[Starburst--driven galactic outflows]{
Starburst--driven galactic outflows
}

\author[Biman B. Nath \& Joseph Silk]{Biman B. Nath$^1$
\& Joseph Silk$^2$\\
1.  Raman Research Institute, Sadashiva Nagar, Bangalore 560080, India\\
2.  Department of Astrophysics, University of Oxford and Beecroft Institute for Particle Astrophysics and Cosmology\\
Denys Wilkinson Building, Keble
Road, OX1 3RH Oxford, UK}

\begin{document}
\maketitle

\begin{abstract}
We propose a model of starburst--driven galactic outflows
whose dynamics depends on both
radiation and thermal pressure. 
Standard models of thermal pressure--driven winds
fail to explain some key observations of outflows at low and high redshift
galaxies. We discuss a scenario in which radiation pressure from massive stars 
in a stellar population
drive a shell of gas and dust. Subsequent supernova (SN) explosions from the
most massive stars in this population then drive stellar ejecta outward in 
a rarefied medium, making it collide with the radiation pressure driven shell.
The collision imparts renewed momentum to the shell, and the resulting
re-acceleration makes the shell Rayleigh-Taylor unstable, fragmenting the shell.
We show that the speed of these ballistic fragments can explain some recently
observed correlations in Lyman break galaxies between wind speed, reddening
and star formation rate.
\end{abstract}


\section{Introduction}
Removal of gas from galaxies in the form of outflows constitute an important
phase in the evolution of galaxies, and an important mechanism of feedback
from galaxies to the surrounding medium. At low redshift, outflows have been
directly observed in many galaxies, in optical to X-rays, and at high redshifts,spectroscopic studies have found many examples of outflows. Outflows have
been ascribed to either the stellar population in a galaxy or active galactic
nuclei, or both (see Veilleux, Cecil, \& Bland-Hawthorn 2005, for a recent review).

In the standard scenario of starburst--driven
galactic outflows, thermal pressure builds up in 
the interstellar medium (ISM) of a galaxy owing to supernovae shocks that
results from formation of stars, and the pressure of the hot gas then 
removes a fraction
of the ISM in the form of an outflow (Veilleux, Gerald, Bland-Hawthorn 2005, 
and references therein). 
The pressure of the hot gas is directly
related to the rate of supernovae, or, equivalently to the star formation
rate.
This scenario, however, has difficulty in explaining some key observations at
both low and high redshifts.

Starburst--driven winds observed in nearby galaxies through NaI absorption
lines show a correlation between the maximum speed of clouds embedded in 
the outflow, and the star formation rate (Martin 2005). This can be explained
by ram pressure--driven clouds embedded in the outflow, except for very large
outflow speeds (Murray, Quataert \& Thompson 2005) that may require
radiation pressure (Veilleux, Gerald, Bland-Hawthorn 2005). 
Martin (2005) also showed
a correlation between maximum cloud speed and the rotation speed of host 
galaxies, which she suggested is an indication of dominant contribution
from radiation pressure.

Murray, Quataert \& Thompson (2005) have discussed the basic dynamics of
a radiatively driven shell of gas and dust. Using these models, Martin (2005)
showed that her data was indicative of a feedback mechanism on the star
formation process, in which the star formation rate was regulated by gas removal
through radiation pressure.

At high redshifts,
Shapley \etal (2003) studied a set of $\sim 800$ Lyman break galaxies
(LBG) at $z\sim 3$, and found that the doppler shift speed
$\Delta v=v_{em}-v_{ab}$
of the  Lyman-$\alpha$ line is 
correlated with the star formation rate (as implied
by the $\mathcal{R}$ luminosity),
as well as with the reddening due to dust. The outflow speed (which is related
to $\Delta v$) was also found to be correlated with the equivalent width
of low ionization absorption lines. Taken together with the correlation
with the reddening, this suggested that most of the reddening and low ionization
absorption occurred in the same region, near the clouds responsible for the
doppler shift.

Furthermore, Ferrara and Ricotti (2006) suggested after studying the
outflow speed ($V_w \sim 0.5 \Delta v$) and the SFR ($\dot{M}$) 
that it was difficult to
explain the relation with thermal pressure--driven winds. The observed
relation is approximately $V_w \propto \dot{M} ^{2/3}$. On the contrary,
in the standard SN--driven wind via the Sedov-Taylor phase of SN shells, one
expects $V_w \propto E^{1/5}$, where $E \propto \dot{M}$ is the rate of 
energy deposited by the SN.

The correlation between reddening and wind speed observed
in LBGs motivates us to seek a model where radiation pressure is a dominant
feature. In the case of a purely radiatively driven shell of gas, the speed
is proportional to the star formation rate (Murray et al. 2005), so it is not
clear whether one can explain the above mentioned correlations with radiation
pressure alone. At the same time, 
it is reasonable to assume that if the pressure
of continuum radiation comes mostly from massive stars (the continuum dust
opacity being larger in the blue than in the red), 
then these stars would also explode
earlier than the rest of the stellar population. It is conceivable that a
radiatively driven wind shell 
is pushed by thermal pressure from SN explosions at a 
later epoch.

Here we study a hybrid model in which first a radiation pressure--driven wind is
formed,  and a shell moves outward. Then, after a few  million years, 
supernova explosions of the most massive stars will 
drive their stellar ejecta through a 
comparatively rarefied medium, and will collide with the radiatively driven
shell. The shell will become vulnerable to fragmentation at this stage from 
Rayleigh-Taylor instability, and the resulting
clouds will move ballistically through the ISM, with the momentum acquired 
from the collision. We show that this model explains the correlation
between outflow speed and SFR. In addition, in this model, the dust in the
cloud, responsible for the dynamics of the radiatively driven shell,
can explain the relation between reddening and outflow speed, as observed in
LBGs.

\section{Wind dynamics}
\subsection{Phase 1: radiation pressure--driven wind}
We first study a radiatively driven shell of gas and dust. We
 assume a point source of radiation, of luminosity $L$, and assume
a spherically symmetric shell of gas at distance $r$. Suppose the
total mass (including dark matter)
contained within this distance is $M(r)$. Setting  $\kappa$
to be  the opacity of matter (dust plus gas) per unit mass of gas (in
cm$^2$ gm$^{-1}$), and ignoring gas pressure inside the cavity of the
shell, the equation of
motion of the gas shell is given by
\begin{equation}
{dV \over dt}=-{G M(r) \over r^2} + {\kappa L \over 4 \pi r^2 c} -{3 V^2 \over
r} \,.
\label{eq:rad}
\end{equation}

In a time--independent case, this equation can be re-written for a wind,
and solved for the wind speed at a given radius $r$, as
(ignoring the last term, which refers to deceleration due to swept--up mass;
see Murray, Quataert \&
Thompson 2005),
\begin{equation}
{1 \over 2} V(r)^2 = \int _{R_0} ^r {G M(r) \over r^2} \Bigl (
{L \over L_{Ed} (r)}
-1 \Bigr ) \,.
\end{equation}
Here $L_{Ed} (r)={4 \pi G M(r) c \over \kappa}$ is the Eddington luminosity relevant for the given
radius $r$ given the mass distribution $M(r)$ and opacity $\kappa$.
The star formation rate corresponding to $L_{Ed}(r)$ is given by
$\dot{M}_\ast \sim L_{Ed}(r)/(0.007 c^2)$. For a singular isothermal
mass distribution with velocity dispersion $\sigma$ ($M(r)=2 \sigma^2 r/G$),
this star formation rate is $\sim 50 \, \sigma_{200}^2 \, (r/10 \, {\rm kpc})$
$M_{\odot}$ yr$^{-1}$, for $\kappa \sim 500$ cm$^2$ gm$^{-1}$ and where
$\sigma= 200 \, \sigma_{200} $ km s$^{-1}$.

\subsection{Dust opacity}
Many observations of galactic outflows indicate the presence of dust grains
in the outflowing gas. Martin (2005) has noted that the absence of NaI emission
indicates absorption by dust, and that cold dust is often found in starburst
galaxies at larger scale heights than those of the stars.
Opacities of order $\kappa \sim 500$ cm$^2$ gm$^{-1}$ can be provided
by dust grains through continuum absorption and scattering (Draine \& Lee
1984).  The absorption opacity of gas mixed with dust
is expected to be larger at shorter wavelengths:
$\kappa \sim 600\hbox{--}100$ cm$^2$ gm$^{-1}$ at $\lambda \sim 
1000\hbox{--}6000$ $\AA$ (Li \& Draine 2001). 
Observations of reddening of local (Calzetti 2001 and references
therein) and high redshift starburst
galaxies (Sanders \& Mirabel 1996), including LBGs (Adelberger \& Steidel
2000) also suggest dust opacities of this order. 
In our calculations described below, we assume dust opacity of
order $\kappa \sim 500$
cm$^2$ g$^{-1}$.

\subsection{Evolution in luminosity}
Consider the case of a single burst of star formation  with an initial
mass function (IMF) $\phi(m)=\phi_1
(m/m_1)^{-(1+x)}$, which is normalized as $M_o=M_o \int m \phi (m)dm$, where
$M_o$ is the total mass of the stellar population. 
The total luminosity of main sequence stars in
 a stellar population evolves as,
\be
L (t) \propto \int _{m_l} ^{m_u} \phi(m) l(m) dm \sim {M_o \phi_1 m_1 l_1
\over \alpha -x} (t/\tau_1)^{{(\alpha -x) \over (1 - \alpha)}} \,,
\label{eq:lum1}
\ee
where $l(m)=l_1 (m/m_1)^{\alpha}$ is the main sequence luminosity of a 
star with mass $m$, and $\alpha \sim 3.5$. The main sequence lifetime
of a star with mass $m_1$ is denoted here by $\tau_1$. Stars on the giant
branch contribute more to the luminosity, and their contribution can be
incorporated by including a multiplicative factor $[1+G(t)]$, where
$G(t)={\alpha -x \over \alpha -1}{
l_g \tau_g \over l_g (m_{tn}) t}$, and  $l_g, \tau_g$ are the giant branch
luminosity and lifetime on the giant branch, and $m_{tn}$ is the turn-off
mass at time $t$ (Tinsley 1980). The ratio $G(t)$ is of order unity
for short wavelengths (although larger for longer wavelengths, where giant
branch stars contribute most), and so we multiply the expression in
equation \ref{eq:lum1} by a factor 2.  

In the case of continuous star formation, with a constant SFR,
it can be shown that the luminosity evolves as
\be
L(t) \approx { [1+G(t)] \dot{M} \phi_1 m_1 l_1
\over \alpha -x} \, (t/\tau_1) ^{{1-\alpha \over 1 -x}} 
\approx
4 \times 10^9 \, L_{\odot} \dot{M}_1  \,t_6^{0.14}\,,
\ee
where $t_6=t/1$ Myr, $\dot{M}_1=\dot{M}/1$ M$_{\odot}$ yr$^{-1}$, 
and where
we have assumed a Salpeter IMF ($x=1.35$), and that $G(t) \sim 1$. 
We use this expression for  luminosity in equation \ref{eq:rad} for
our calculations below.

\subsection{Phase 2: supernova explosions}
Next,
we consider the evolution of the stellar ejecta from supernova explosions
of the most massive stars in this population. The ejecta will move through
a rarefied medium in the interior of the radiatively driven shell, since
a large fraction of gas is expected to be swept out earlier by radiation
pressure.  We
consider  supernovae resulting from the early stages of a starburst,
within an interval $t_{sn} \sim 3$ Myr, which is the main sequence 
lifetime of stars with $M>15 M_{\odot}$. Assuming a Salpeter IMF, one gets
a total mass of ejecta (ignoring the remnant masses), as $M_{ej}\sim 0.1
\dot{M} t_{sn}$, considering supernovae for $M>15 M_{\odot}$, and a lower and
upper mass cut-offs of $0.1$ and $125$ M$_{\odot}$. 
The total energy output is related to the number of SN,
$\sim 0.003 \, \dot{M} \, t_{sn}$, again assuming a Salpeter IMF and supernovae
for $M>15 M_{\odot}$. We assume
an energy output of $10^{51}$ ergs per supernova, so that the total
energy output is $E_{ej}=10^{51} {\rm erg} \, \times 0.003 \,\dot{M} \, 
t_{sn}$. 

We also assume that the evacuated and rarefied
 ISM in the interior of the radiatively
driven shell  has a particle density $n=n_r$. In the Sedov-Taylor phase (which
is valid since the ejecta travel with high speed until  colliding with the
radiatively driven shell, and since the shell does not sweep up much 
matter in the
rarefied medium to slow it down), we have for the evolution of the
outer shell of the ejecta (e.g., Shull \& Silk 1979),
\bea
R_{ej}&\approx& 9.89 \times 10^{-2} \, {\rm kpc} \, 
E_{ej, 51}^{0.2} \, n_{r,1} ^{-0.2}
\, t_6^{0.4} \nonumber\\
V_{ej}&\approx& 38.69  \, {\rm km} {\rm s}^{-1} \, 
E_{ej, 51}^{0.2} \, n_{r,1} ^{-0.2}
\, t_6^{-0.6} \,,
\label{eq:sn}
\eea 
where $E{ej,51}=E_{ej}/10^{51}$ erg, and $n_{r,1}=n_r/1$ cm$^{-3}$.

\subsection{Re-acceleration of the shell and growth of instability} 
The radiatively driven shell is first accelerated for a short period, and
then soon begins to decelerate as the shell radius $r$ increases and the
radiation pressure (which scales as $r^{-2}$) falls. At the same time,
the shell of ejecta from subsequent SN events
moves swiftly through the rarefied medium inside the radiatively
driven shell, and catches up with it at a time $t_{coll}$, when they collide.
 The speed
of the ejecta is in general  much larger than the shell speed, and although the
ejecta mass is very small, the momentum imparted to the shell by the colliding 
ejecta is large. Therefore, the shell is carried forward by the increased
momentum from the ejecta coming up behind it. The resulting speed of the
shell is given by
\be
v= (M_{ej} V_{sn}+ M_{shell} V_{shell})/(M_{ej}+M_{shell}) \,,
\ee
where $M_{shell}$ and $V_{shell}$ are the swept--up mass in the shell and the
speed of the shell at the time of collision, and $V_{sn}$ is the speed of the
ejecta at that time.

After the collision, the wind shell is re-accelerated, and we have the
case of a heavy fluid (in the shell) above a rarefied medium, both
being acted upon by an effective gravity
resulting from the acceleration of the shell. The shell material is therefore
subject to Rayleigh-Taylor (RT) instabilities, and is liable to fragmentation.

Even before the collision, the expanding shell is prone to dynamical
and gravitational instabilities. Vishniac (1983) estimated the 
shortest growth time for gravitational instability to be given by
$t_g\sim c_s/\pi G \sigma$, where $\sigma$ is the surface mass density.
A lower limit on the growth time for instability is obtained by assuming a 
cool shell with $T\sim 10^4$ K. 

After the collision, the shell is re-accelerated, and becomes Rayleigh-Taylor
unstable.
The effective gravitational acceleration at this time is $g_{eff} \sim v^2/r$.
The time scale of Rayleigh-Taylor instability for clouds
of size $\lambda=2 \pi/k$ is given by,
\be
\tau_{RT} \sim (g_{eff} k)^{-1/2} \sim (\lambda/ 2 \pi g_{eff})^{1/2} \,,
\label{eq:rt}
\ee
where $k$ is the wavenumber.

\section{Results}
We follow the evolution of the radiatively driven shell (eqn \ref{eq:rad})
numerically and track the trajectory of SN ejecta (eqn \ref{eq:sn}), and
determine the time of collision ($t_{coll}$) between the two. 
For the dark matter profile, we have assumed the galaxies to be of total mass
$M=10^{12} M_{\odot}$, at $z=3$, and assumed a Navarro-Frenk-White halo
profile, with a concentration parameter $c=3$ (Navarro, Frenk, White 1996).
Our assumption is motivated by the estimate made by Adelberger \etal (2005)
from the clustering of LBGs and comparison with N-body simulations. The 
median halo mass of LBGs at $z\sim 3$ is estimated to be $10^{11.86\pm 0.3}$ 
M$_{\odot}$. Although there is considerable uncertainty in the dynamical
mass estimate of LBGs (see
Weatherley \& Warren 2005), our results are not sensitive to the gravitational
deceleration of the shell by dark matter, and therefore, to the assumption of
total mass and its profile.

Figure 1 shows the evolution of the shell radius (solid line, top 
panel) and speed (solid line, bottom panel) with time,
for a star formation rate of $\dot{M}=100$ $M_{\odot}$ yr$^{-1}$, a continuum
dust opacity $\kappa=500$ cm$^2$ gm$^{-1}$. The particle density in the
undisturbed ISM (that the radiatively driven shell sweeps through) is assumed
to be $n_g=5$ cm$^{-3}$, and the density inside the shell is assumed to be
$n_r=0.01$ cm$^{-3}$. It is seen that the shell accelerates for an initial
brief period ($\ll 1$ Myr), after which it continuously decelerates, mostly
owing to the decrease in radiation pressure with distance, and to some 
extent, to the growing weight of the swept--up mass in the shell.

\begin{figure}
\centerline{
\epsfxsize=0.3\textwidth
\epsfbox{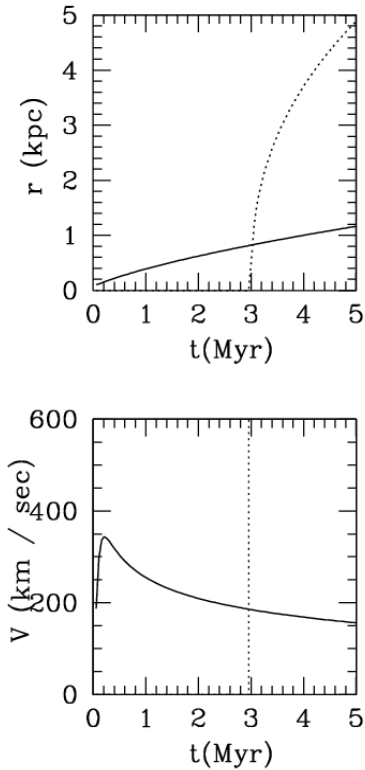}
}
{\vskip-3mm}
\caption{
The top panel plots the shell radius (kpc)  with time (Myr)
(solid line),
for a star formation rate of 100 $M_{\odot}$ per year, with ambient gas
density $n_g=5$ per cc, and for opacity $\kappa=500$ cm$^2$ gm$^{-1}$.
The dotted line shows the evolution of the hot gas bubble, assuming a
density inside the shell $n_r=0.01$ per cc.
The bottom panel shows the corresponding speed of the shell (solid line)
and the hot bubble (dotted line).
}
\end{figure}

The dotted lines in Figure 1 show the corresponding evolution for the
stellar ejecta, which we assumed to expand outward
after a period of $t_{sn} \sim 3$ Myr. 
The rarefied medium inside the radiatively
driven shell helps the ejecta move with high speed, and catch up with the
shell within a short time. 

Figure 2 plots the (logarithm of the) 
ratio of the shortest time--scale for instability
to the age of the shell, as a function of time. The dotted line shows the
shortest time--scale for gravitational instability, assuming a cool shell with
gas at $T\sim 10^4$ K. The surface mass density is calculated from the swept--up mass  within $r$, the shell radius. The figure shows that gravitationally
unstable modes takes $\gg 10$ Myr to grow, although the time--scale for growth steadily
decreases with time, due to the increasing surface mass density.

After the collision, the shortest time--scale
for RT instability is calculated by using equation \ref{eq:rt} for instabilities
of wavelength $\sim 2\pi r$, where $r$ is the shell radius. This is the largest
wavelength for instabilities in the shell, and therefore gives the time--scale
for the fastest growing mode.
This time--scale  at the time of collision is shown as a point.
The effective gravitational acceleration at this time is $g_{eff} \sim v^2/r
\sim 10^{-7}$
cm s$^{-2}$, for 
$\dot{M} \sim 100$ $M_{\odot}$ yr$^{-1}$, making the shortest time--scale for
instability growth comparable to the age of the shell (a few Myr). The
re-acceleration of the shell, therefore, makes it prone to RT instabilities. 
This
is likely to feed gravitational instabilities on scales that are gravitationally
stable at this stage.

\begin{figure}
\centerline{
\epsfxsize=0.3\textwidth
\epsfbox{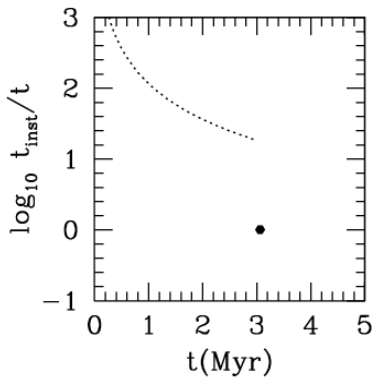}
}
{\vskip-3mm}
\caption{
The ratio between the instability growth
time--scale and age of the shell (in logarithmic units) is plotted with time.
The dotted line shows the
shortest time--scale for gravitational instability, assuming a shell temperature
of $10^4$ K. For most of the time span, the shell decelerates and so there
is no RT instability, but after the collision, it is accelerated, and the
ratio of RT timescale and age of the shell is shown as a point at the time
of collision.
}
\end{figure}

The size of the resulting fragments will be $1\hbox{--}1.5$ kpc, 
the shell radius at this
point of time, as plotted in Figure 3 as a function of the outflow
speed for the assumed
values of the parameters.
We then identify the speed of the cloud after collision as the outflow speed, 
$V_w$. We can then compare our results with data from Shapley \etal (2003), for
different assumptions of parameter values. The results of the present model
are insensitive to the launching radius and initial speed of the shell, and
depends on the assumed values of dust opacity ($\kappa$), the densities of
gas outside ($n_g$) and inside ($n_r$) the shell.
We found that the best case scenario needs $\kappa=500$ cm$^2$ gm$^{-1}$,
$n_g=5$ cm$^{-3}$,
and $n_r=0.01$ cm$^{-3}$.

\begin{figure}
\centerline{
\epsfxsize=0.3\textwidth
\epsfbox{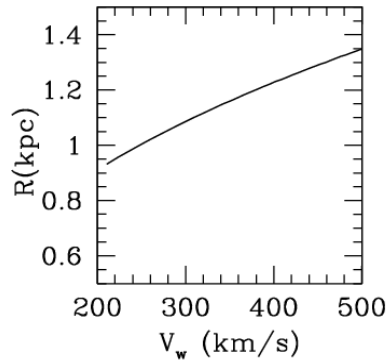}
}
{\vskip-3mm}
\caption{
The shell radius at $t_{coll}$ as a function of outflow speed (in km/s)
for the
same set of parameters as in Figure 1.
}
\end{figure}

The top panel of
Figure 4 plots the resulting outflow speed with SFR for these parameter values ,
where we show the result of our calculation with a solid curve, along with
data points from Shapley \etal (2003) (their Table 3, with $h=0.7$).
The dashed line in the top panel shows the slope of the
expected relation between
outflow speed and SFR (scaled arbitrarily)
in the case of a shell driven by hot gas
($V_w \propto E^{1/5}$).

Another natural prediction of this scenario is a correlation between
outflow speed and reddening due to dust absorption.  The outflow speed is
determined by the the combined momentum of the radiative shell and the
ejecta, which is related to the total mass in the shell, and therefore, to 
the amount of dust in the shell. We can quantify this
correlation by estimating the amount of reddening using the column density
of material in the shell.
The bottom panel of Figure 4 shows
the expected reddening (E(B-V)) as a function of the cloud speed for three
different choices of the reddening versus column density relation. 
The solid line
at the bottom of the figure uses the relation
between hydrogen column density and reddening as found in the Small Magellanic
Cloud (SMC)
($N_{H}/E(B-V)=4.4 \pm 0.7 \times 10^{22}$ cm$^2$ mag$^{-1}$;
Bouchet et al 1985), with the dotted lines showing the
$1-\sigma$ spread. The short--dashed line uses the calibration from
observations of the Large Magellanic Cloud (LMC; $N_{H}/E(B-V)=2 \pm 0.5 
\times 10^{22}$ cm$^2$ mag$^{-1}$;
Koornneef et al 1982), and the long--dashed line uses the Galactic ISM data 
($N_{H}/E(B-V)=4.93 \pm 0.28 
\times 10^{21}$ cm$^2$ mag$^{-1}$;
Diplas \& Savage 1994).

\begin{figure}
\centerline{
\epsfxsize=0.3\textwidth
\epsfbox{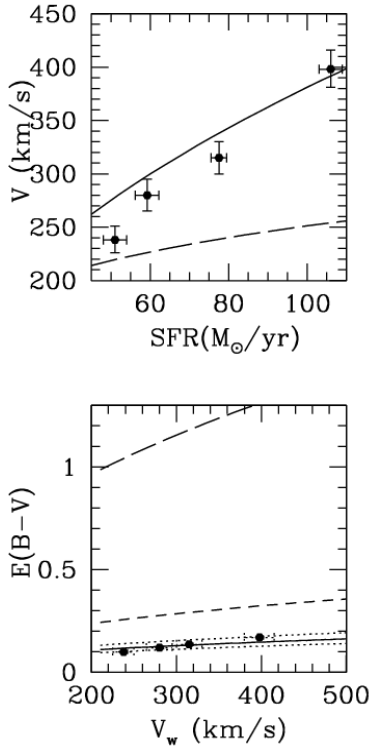}
}
{\vskip-3mm}
\caption{
The top panel plots the outflow speed with SFR with a solid
curve. The dashed curve is the expected result from a thermal 
pressure--driven wind. The bottom panel plots the reddening with outflow speed
using calibration of column density and E(B-V) from SMC (solid line),
LMC (short--dashed line) and Galactic ISM (long--dashed line).
 The dotted lines show the 1-$\sigma$ error margin for the calibration
for SMC type dust.}
\end{figure}

\section{Discussion}
The top panel of Figure 4 shows that the scenario described here,
along with suitable assumptions of the values of key parameters
($\kappa=500$ cm$^2$ gm$^{-1}$,
$n_g=5$ cm$^{-3}$,
and $n_r=0.01$ cm$^{-3}$), can explain the observations of outflows speed
and SFR observed
in LBGs. 


While the calibration of column density and reddening from SMC data explains
the data of reddening and outflow speed in LBGs from Shapley \etal (2003), the
expected reddening from Galactic ISM calibration is much larger. It should
be kept in mind that dust grains are destroyed in ISM shocks by a
variety of processes, from energetic grain-grain collisions and from
collisions between gas ions and dust grains. Calculations show that large
grains (with sizes larger than $\sim 0.05$ $\mu$) are destroyed in shocks
with speed larger than $\sim 200$ km/s (Jones \etal 1997). In the present
case, dust grains embedded in the radiatively driven shell coast with
a speed $\sim 200$ km s$^{-1}$ for a few Myr before being hit by stellar
ejecta (see bottom panel of Figure 1). Afterwards, the
fragmented clouds traverse the rest of the ISM at high speed ($v\sim 300\hbox
{--}400$ km s$^{-1}$) and leave the host galaxy within a few Myr, because
the typical half-light radius of LBGs is of order $\sim 1$ kpc (see below), 
comparable
to the distance where clouds form in the present scenario.
Dust grains in the cloud will therefore be destroyed to a certain extent,
especially the larger grains, and the dashed and dotted
lines in Figure 4 (bottom panel),
corresponding to Galactic and LMC-type dust, will come  closer to the
LBG data.

Another interesting possibility is that due to the destruction of large
grains, the grain size distribution in these clouds will be dominated by
smaller ($\le 500 \, \AA$) grains, and so the total surface area of grains will
increase, thereby increasing the reddening for shorter
wavelength ultraviolet continuum. The SMC-type extinction curve shows
more reddening at wavelengths shorter than the $2175 \AA$ feature
 than the Milky Way extinction curve (e.g., Pei 1992), and so the solid
line in Figure 4, corresponding to SMC--type dust, is likely to show increased
reddening for larger outflow speeds, thereby bringing the predicted curve
closer to the data points in Shapley \etal (2003). At the same time, the
smaller grain size would show less reddening in the optical. Therefore, the
frequency dependence in reddening for outflows in LBGs will likely change with
redshift, as the rest frame wavelength for observed optical radiation changes
with redshift.


An important caveat arises from the fact that in comparing with the data,
we have identified the outflow speed
with $\Delta V/2$, where $\Delta V=\vert V_{em}-V_{abs}\vert$. Adelberger \etal (2005) 
have shown that there are
asymmetries between $V_{em}-V_{neb}$ and $V_{abs}-V_{neb}$, where
$V_{neb}$ is the speed of the ISM, indicating that outflow speed can significantly
differ from $\Delta V/2$, which we have assumed for simplicity.

The scenario discussed here has several implications. Firstly, the shell fragments
are generated at radii $\sim 1\hbox{1.5}$ kpc (Figure 3), 
comparable to the half-light
radii of LBGs. Bouwens \etal (2004) estimated the mean half-light size of LBGs
at $z\sim 3$ as $\sim 1.3$ kpc. Therefore, the fragments are likely 
to penetrate 
the rest of the
galactic ISM and deposit metals into the intergalactic medium (IGM), 
instead of falling back on to the host
galaxies. Galactic outflows have been invoked to explain the metallicity of 
low column density Lyman-$\alpha$ absorbers in the  IGM, and several 
authors have expressed concern that this process of metal enrichment through
outflows might (1) increase the number of absorption lines more than necessary
for consistency with models of structure formation, and (2) dynamically disturb the
absorption systems. It has been suggested that these problems could be overcome if
the wind shell fragments, so that the fragments can coast to large distances (e.g.,
Theuns, Mo \& Schaye 2001). (Recent detailed numerical studies, however, find that
galactic winds may not pose a problem for the interpretation of Lyman-$\alpha$
absorption systems as was previously thought (Cen \etal 2005).) We note that
galactic wind--driven shells  in the scenario discussed here
undergo fragmentation even while traversing the
ISM, and the fragmented shell is likely to spread metals
in the IGM in an inhomogeneous manner, and not dynamically disturb the 
gaseous filaments responsible for Ly-$\alpha$ absorption systems.

\section{Acknowledgement}
We thank the referee for valuable comments.


\begin{thebibliography}{}
\bibitem[]{} Adelberger, K. L., \& Steidel, C. C. 2000, ApJ, 544, 218
\bibitem[]{} Adelberger, K. L., Shapley, A. E., Steidel, C. C., Pettini, M.
Erb, D. K. \& Reddey, N. A. 2005, ApJ, 629, 636
\bibitem[]{} Adelberger, K. L., Steidel, C. C., Pettini, M., Shapley, A. E.,
Reddey, N. A. \& Erb, D. K. 2005, ApJ, 629, 636
\bibitem[]{} Bouwens, R. J., Illingworth, G. D., Blakeslee, J. P., Broadhurst, T. J.,
\& Franx, M. 2004, ApJ, 611, L1
\bibitem[]{} Calzetti, D. 2001, PASP, 113, 1449
\bibitem[]{} Calzetti, D., Bohlin, R. C., Kinney, A. L., Storchi-Bergmann, T.,
\& Heckman, T. M. 1995, ApJ, 443, 136
\bibitem[]{} Cen, R., Nagamine, K. \& Ostriker, J. P. 2005, ApJ, 635, 86
\bibitem[]{} Diplas, A., Savage, B. D. 1994, ApJS, 93, 211
\bibitem[]{} Draine, B. \& Lee, H. M. 1984, ApJ, 285, 89
\bibitem[]{} Ferrara, A. \& Ricotti, M. 2006, MNRAS, 373, 571
\bibitem[]{} Jones, A. P., Tielens, A. G. G. M., Hollenbach, D. J., McKee, C. F.
  1997, AIPC, 402, 595
\bibitem[]{} Koornneef, J. 1982, A\&A, 107, 247
\bibitem[]{} Kozasa, T., Hasegawa, H., \& Nomoto, K. 1989, ApJ, 344, 325
\bibitem[]{} Martin, C. L. 2005, ApJ, 621, 227  
\bibitem[]{} Murray, N., Quataert, Q. \& Thompson, T. A. 2005, ApJ, 618, 569
\bibitem[]{nfw96} Navarro, J. F., Frenk, C. S., White, S. D. M. 1996, ApJ,
462, 563
\bibitem[]{} Nozawa, T., Kozasa, T., Umeda, H., Maeda, K., \& Nomoto, K. 2003,
ApJ, 598, 785
\bibitem[]{} Pei, Y. C. 1992, ApJ, 395, 130
\bibitem[]{} Sanders, D. B., \& Mirabel, I. F. 1996, ARA\&A, 34, 749
\bibitem[]{} Shapley, A. E., Steidel, C. C., Pettini, M. \& Adelberger K. L.
2003, ApJ, 588, 65
\bibitem[]{} Shull, J. M., \& Silk, J. 1979, ApJ, 234, 427
\bibitem[]{} Theuns, T., Mo, H. J. \& Schaye, J. 2001, MNRAS, 321, 450
\bibitem[]{} Tinsley, B. M. 1980, FCPh, 5, 287 
\bibitem[]{} Todini, P., \& Ferrara, A. 2001, MNRAS, 325, 726
\bibitem[]{} Weatherley, S. J. \& Warren, S. J. 2005, MNRAS, 363, L6
\end{thebibliography}
\end{document}